\documentstyle[epsfig,A4,12pt]{article}
\begin{document}
\thispagestyle{empty}      

 \begin{center}   
\vskip 2cm
{\Large\bf { On the similarity of the parton distributions in nuclei
with more than three nucleons  }}
\vskip 0.8cm
{\large  G.I. Smirnov.}\\
\vskip 0.8cm
Joint Institute for Nuclear Research, 141980 Dubna, Russia.\\

Current address: CERN, PPE, 1211 Geneva, Switzerland\\
\begin{minipage}[h]{6.3cm}

email: G.Smirnov@cern.ch\\[0.3cm]
\end{minipage}
 
{\em( Contributed paper {\bf{ pa06-016}} to the ICHEP'96, Warsaw )}

\vskip 0.8cm

\begin{abstract}
  It is shown that the latest results from NMC and E665 on
the $F_2^A(x)/F_2^{\rm D}(x)$ obtained in the shadowing region, bring
new evidence of the universal $A$ -- dependence of distorsions
of a free nucleon structure function by nuclear medium. The universality holds
in the entire $x$ -range and can be explained as a saturation of the
distortions of the parton distributions in a four-nucleon system.
\end{abstract}
\end{center}
%
%
The effects of the distortion of a free-nucleon structure by
a nuclear medium are usually observed as a deviation from 
unity of the ratio $r^A(x) \equiv F_2^A(x)/F_2^{\rm D}(x)$,
where $F_2^A(x)$ and $F_2^{\rm D}(x)$ are the structure 
functions  per nucleon measured in a nucleus of mass $A$
and a deuteron, respectively.

Below we compare the $A$ dependence of  distortions found
in  the analysis of recent data collected from the DIS of muons on nuclei in
the range 10$^{-3}$ $<x <$ 0.7 by the NMC (CERN)~\cite{ama95,arn95}
 and E665 (Fermilab)~\cite{ad95} collaborations with the distortions 
which we obtain from the analysis of SLAC and BCDMS
data  obtained in the range of $x >$ 0.2~\cite{gomez}.
 
 We consider  structure function distortions as independent of
the  $Q^2$ at which $r^A(x)$
is measured. This is justified by  conclusions about the
$Q^2$ independence of $r^A$ in the range  0.2 GeV$^2$
$< Q^2 <$ 200 GeV$^2$ (c.f. Refs.~\cite{ama95}--\cite{bari}).
 
  In Refs.~\cite{sm94,sm95} it was found that the $x$ dependence of 
$r^A(x)$ can be factorized into three parts 
in the region  0.001 $< x <$ 0.7, in
accordance with the differences in the $r^A(x)$ behaviour
found in the three intervals of the considered range ---
namely  the (1) shadowing,
(2) anti-shadowing and (3) EMC effect regions:
\begin{equation}
r^A(x) \equiv F_2^A(x) / F_2^{\rm D}(x) ~=~ x^{m_1} (1+m_2) (1- m_3 x) .
\label{smir}
\end{equation}
\noindent The parameters $m_{i}$, $i$= 1 -- 3,
 can be treated as  the
distortion magnitude of the nucleon structure function
introduced for each interval. There are two physical reasons for
parametrizing $r^A(x)$ in the form of Eq. (\ref{smir}).
First, as was shown in Ref.~\cite{marti}, the nucleon structure
function behaves as $F_2(x) \sim x^{- \lambda}$
in the range of small $x$, which is motivated by BFKL dynamics.
Hence,  combinations such as  $F_2^A(x) / F_2^{\rm D}(x)$ should obey
a power law as well. Second, the parameters $m_2$ and $m_3$
enter Eq. (\ref{smir}) in a manner similar to the suggestion
of Ref.~\cite{rep160}, whereby local nuclear density is related
 to the deviation of $r^A(x)$ from unity in the range $x >$ 0.3.

\vspace{-0.7cm}
\begin{figure}[t]
\begin{minipage}[t]{0.38 \linewidth}
\begin{center}
\mbox{\epsfysize=\hsize\epsffile{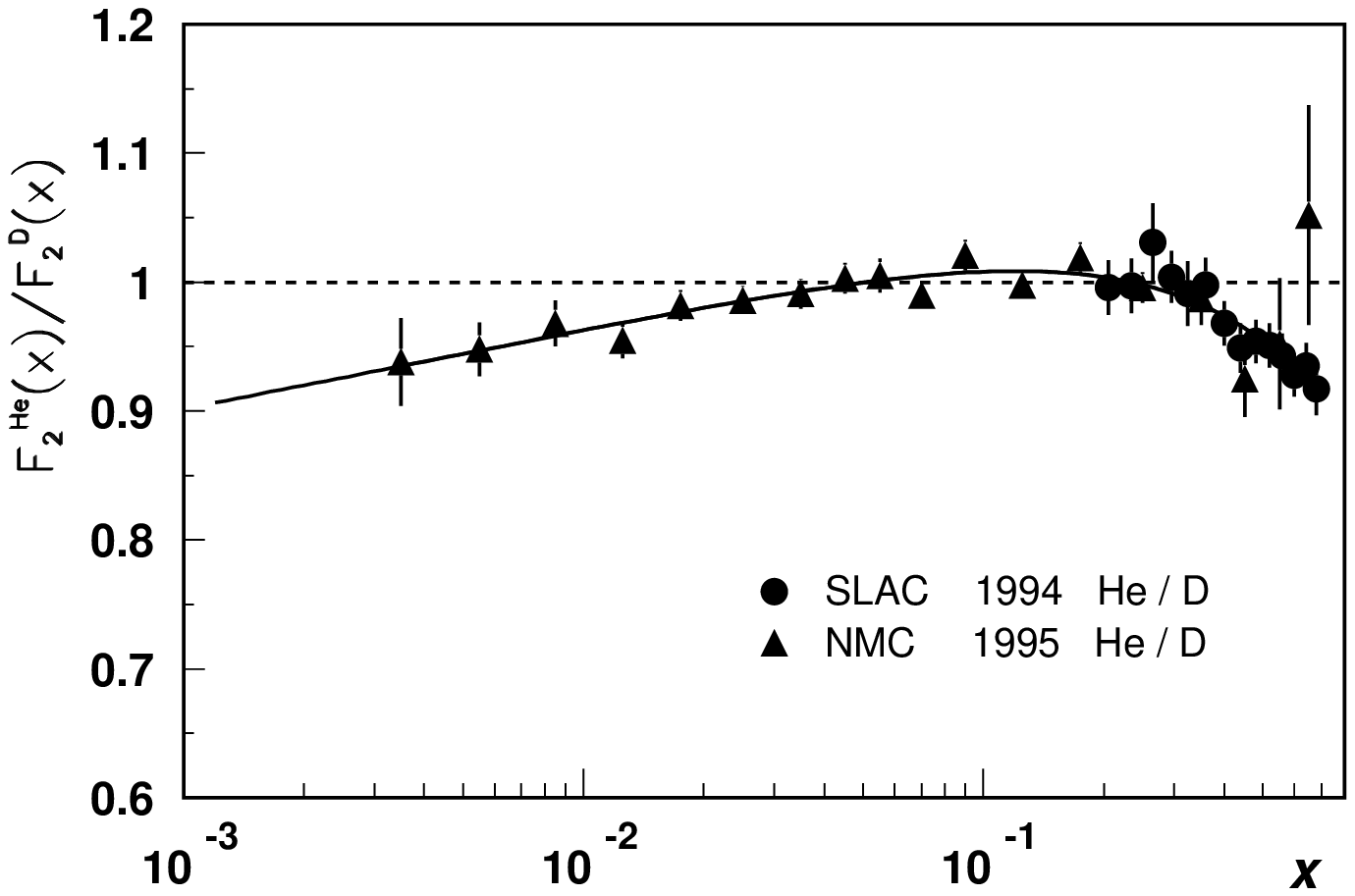}}
\end{center}
\end{minipage}
\hspace{2.2cm}
\begin{minipage}[t]{0.38 \linewidth}
\begin{center}
\mbox{\epsfysize=\hsize\epsffile{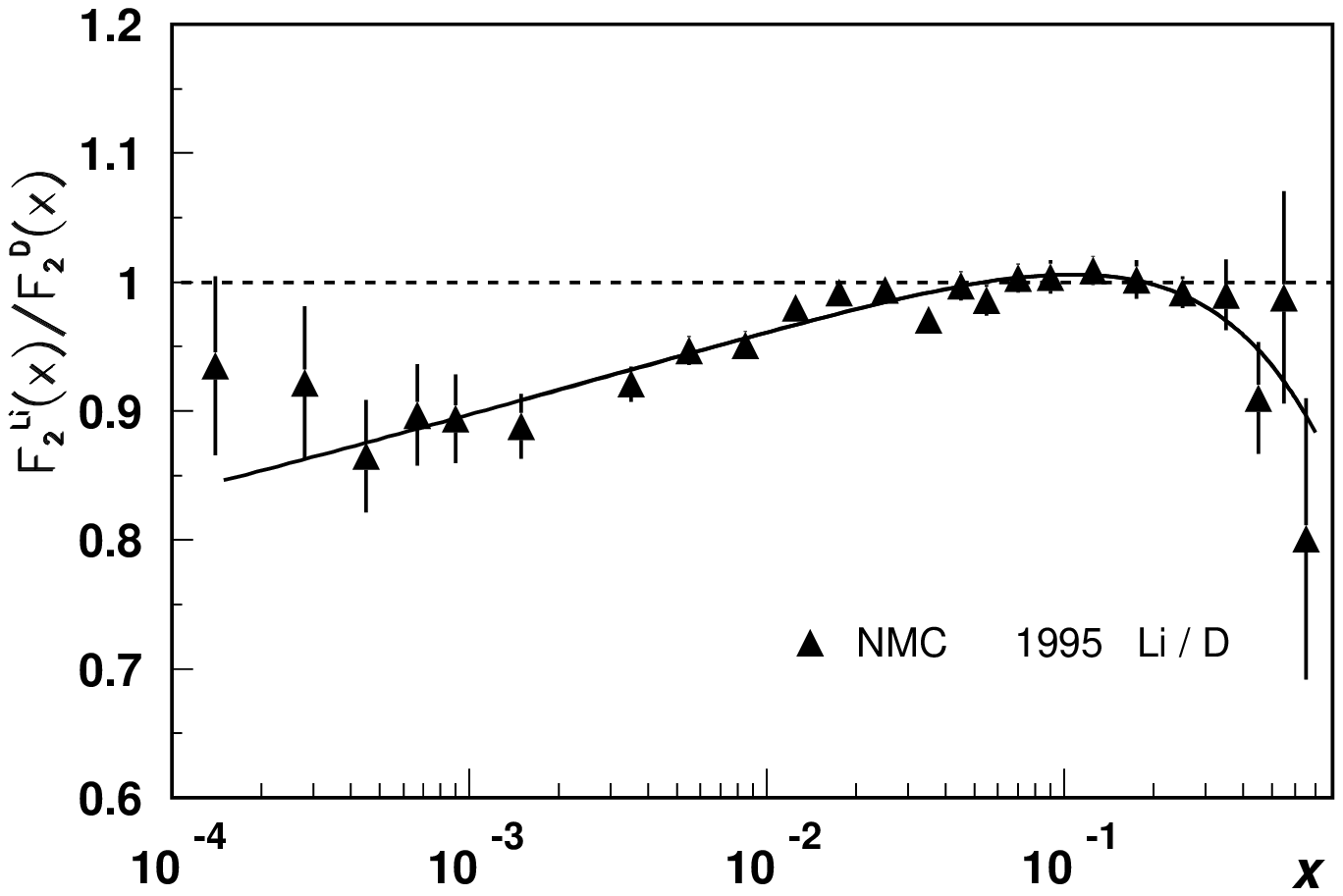}}
\end{center}
\end{minipage}
\vspace{-1.2cm}

\begin{minipage}[t]{0.38 \linewidth}
\begin{center}
\mbox{\epsfysize=\hsize\epsffile{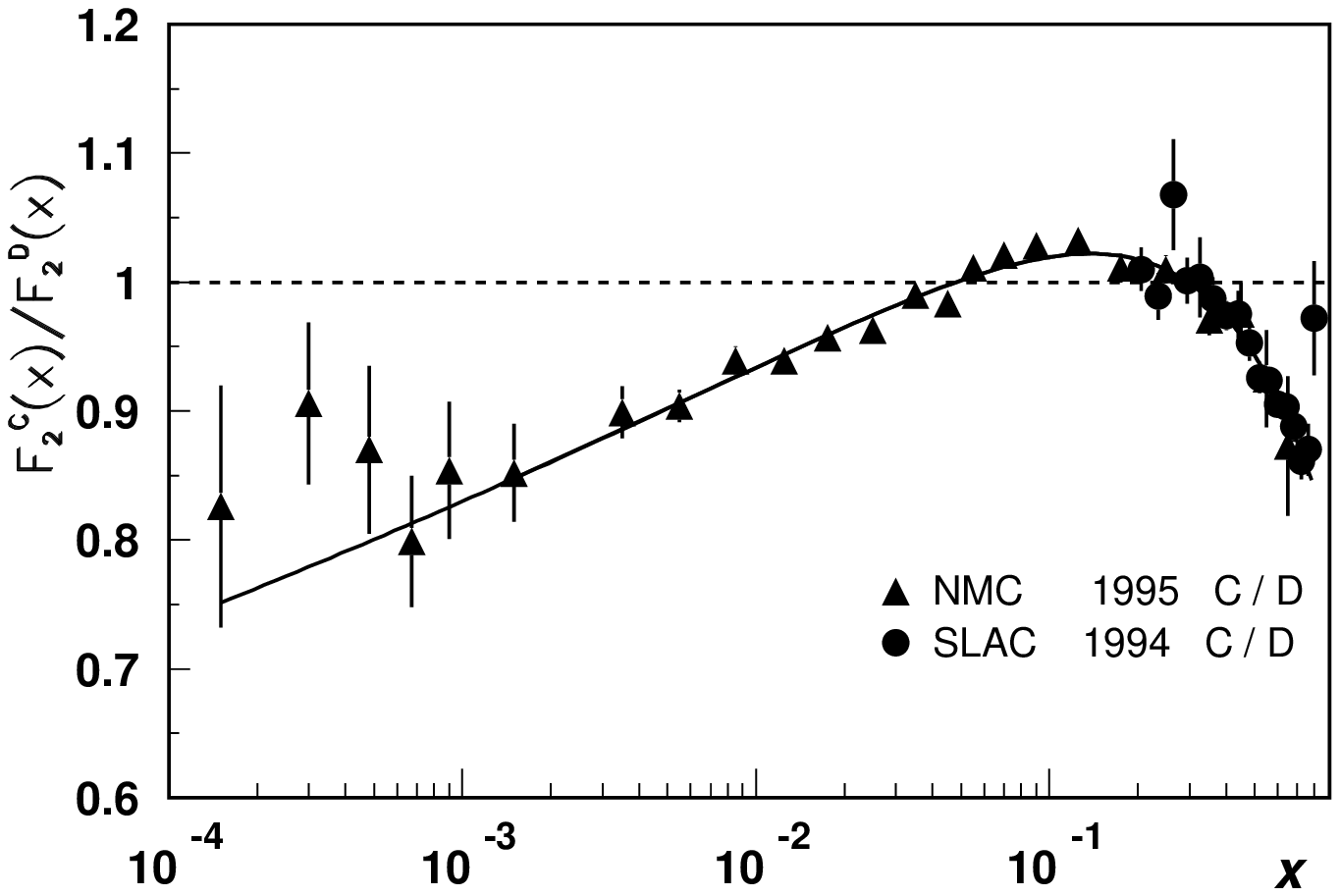}}
\end{center}
\end{minipage}
\hspace{2.2cm}
\begin{minipage}[t]{0.38 \linewidth}
\begin{center}
\mbox{\epsfysize=\hsize\epsffile{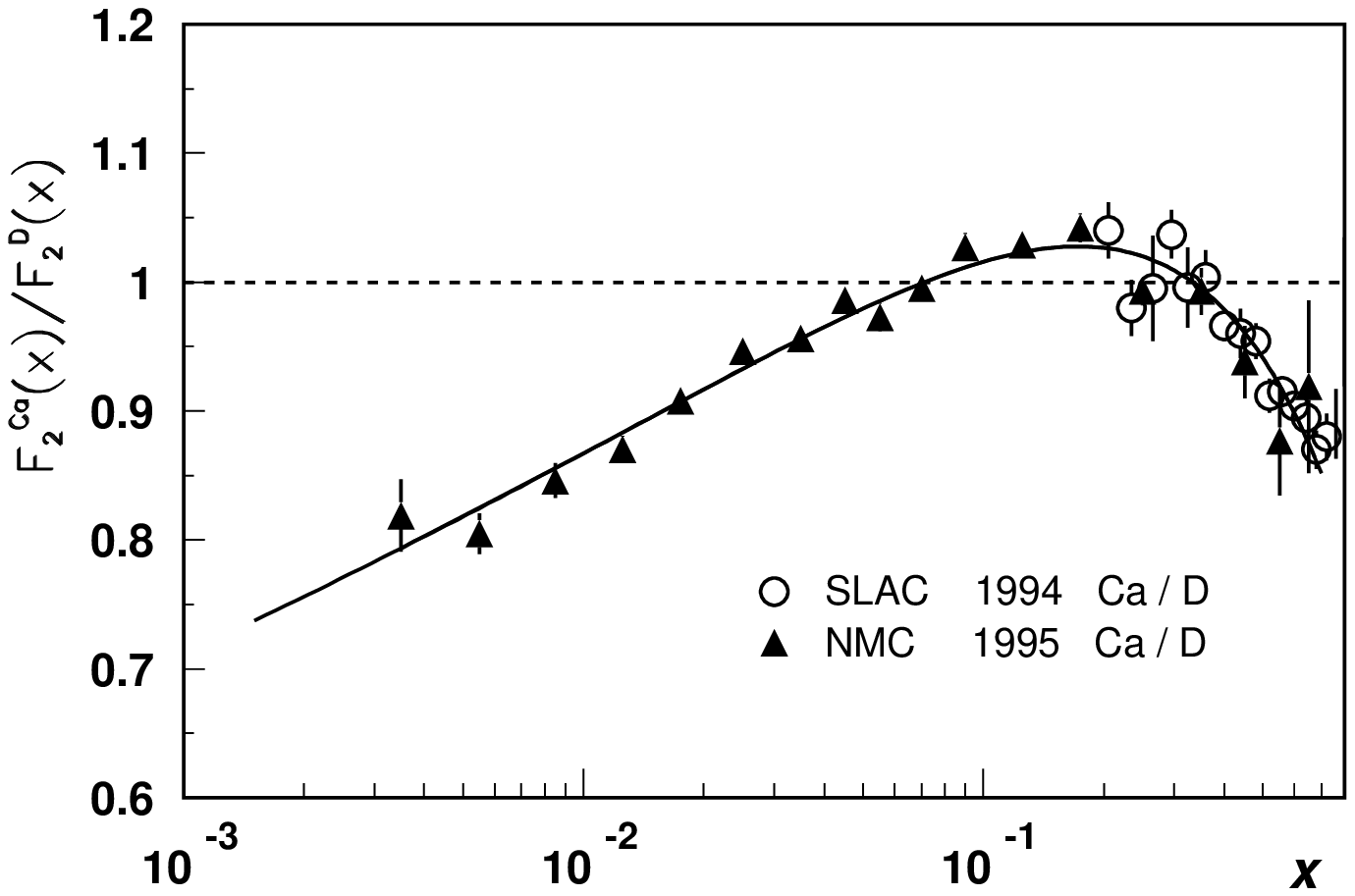}}
\end{center}
\end{minipage}

\caption{ 
 The results of the fit with Eq.(2) of the $F_2^A/F_2^{\rm D}$
measured on $^4$He [3,6], Li [4], C~[4,6] and Ca~[3,6] in the range $x<$ 0.7 .}
\end{figure}

The parameters $m_i$ were determined by fitting $r^A(x)$,
measured on seven nuclear targets --- He~\cite{ama95,gomez},
Li~\cite{arn95}, C~\cite{arn95,gomez}, Ca~\cite{ama95,gomez},
Xe~\cite{xe92}, Cu~\cite{copper} and Pb~\cite{ad95} ---
 with Eq. (\ref{smir}).
We used in the fit the total experimental error determined
by adding statistical and systematic errors at each point
in quadrature. For each of seven nuclei,  good agreement
($\chi ^2 /$d.o.f.$ \leq$ 1) with Eq. (\ref{smir}) was found,
thus proving  that the characteristic pattern   of the 
structure function modifications,  well described
for the helium nucleus by Eq. (\ref{smir}), 
remains unchanged for heavier nuclei. We consider this 
a manifestation of the universality of the $x$ dependence
of the distortions of the free-nucleon structure function
in a nuclear environment.

        The results of the fit are given by the solid line
in Fig. 1 for $^4$He, Li, C and Ca nuclei. 
  The experiment in the SLAC electron beam~\cite{gomez} used a larger
set of nuclear targets. The data however belong to a limited $x$-range, 
0.2 $<x<$ 0.9,  that does not allow to study $r^A(x)$
in the shadowing and anti-shadowing regions. On the other hand they are the
only data, which can be used for the study of the distortions of nucleon
structure functions in the range $x>$ 0.7. In order to perform such analysis we 
approximate the data of Refs.~\cite{gomez,bari} with Eq. (\ref{smir}) modified
by introducing one more parameter ${m'}_4$:
\begin{equation}
r^A(x) \equiv F_2^A(x) / F_2^{\rm D}(x) ~=~ x^{{m'}_1} (1+{m'}_2) (1- {m'}_3 x)
{exp(- ({m'}_4 x)^2) \over (1-x)^{{m'}_4}}~.
\label{smir1}
\end{equation}

\begin{figure}[t]
\begin{minipage}[t]{0.35 \linewidth}
\begin{center}
\mbox{\epsfysize=\hsize\epsffile{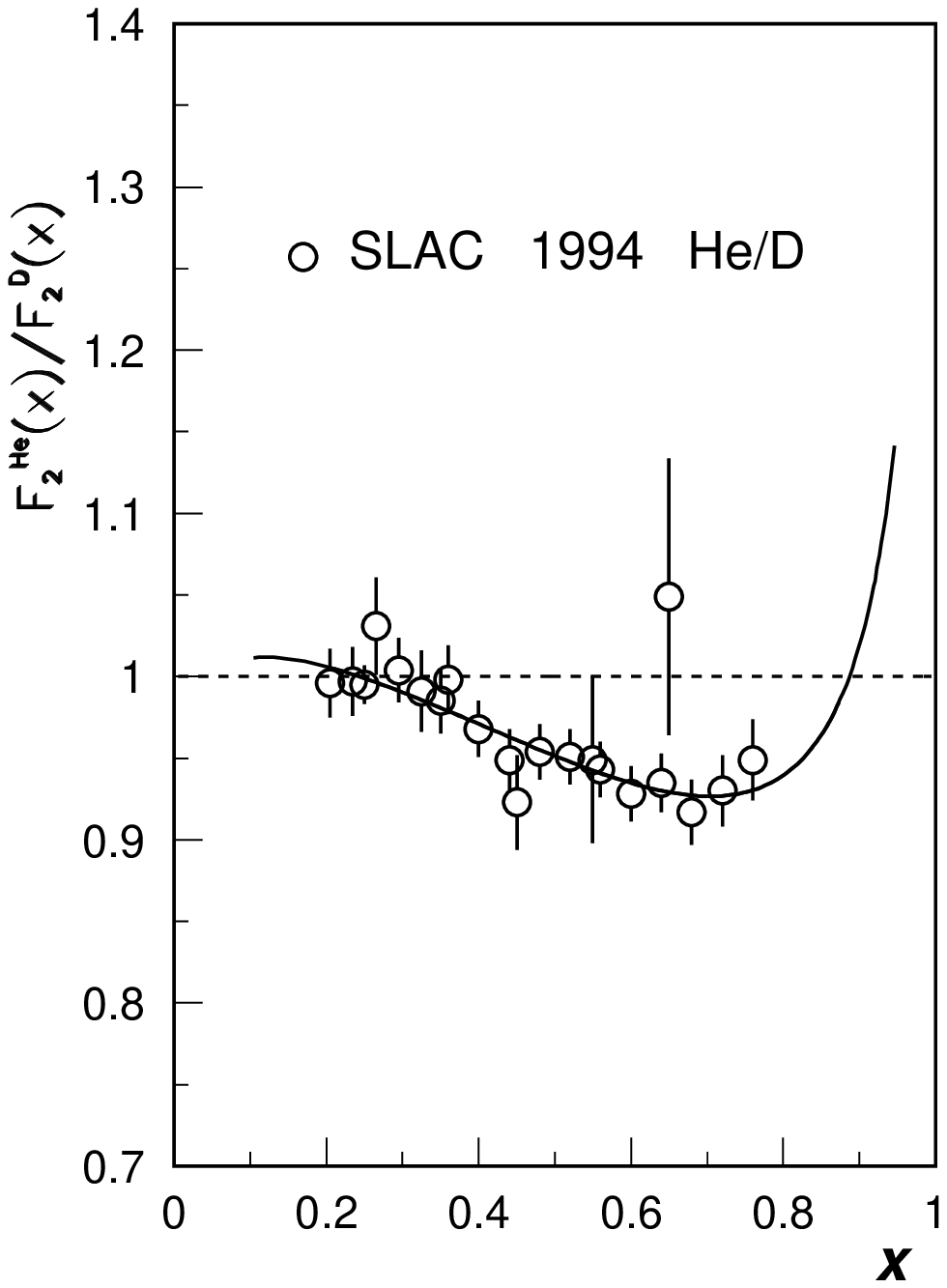}}
\end{center}
\end{minipage}
\hspace{-2.4cm}
\begin{minipage}[t]{0.35 \linewidth}
\begin{center}
\mbox{\epsfysize=\hsize\epsffile{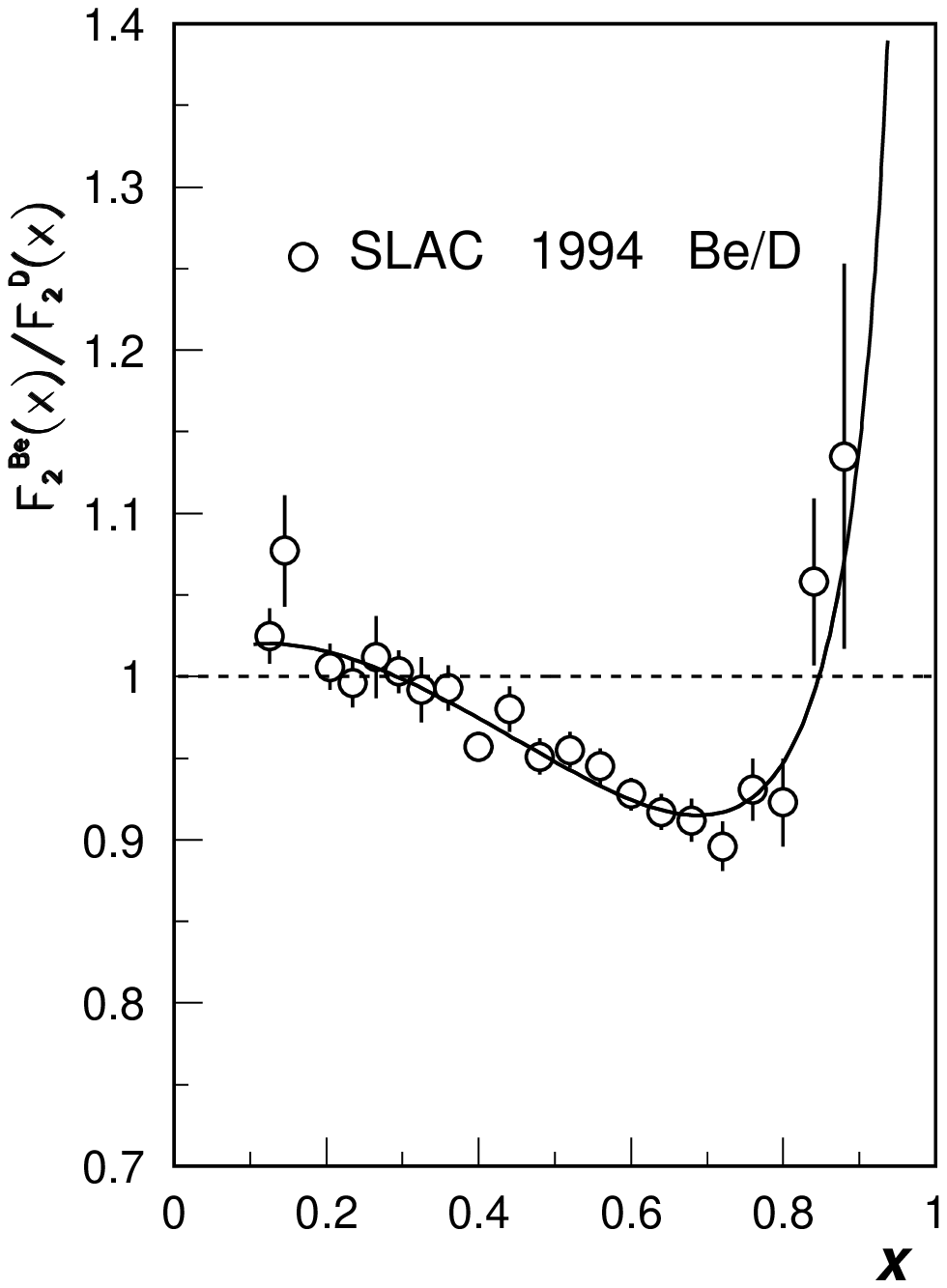}}
\end{center}
\end{minipage}
\hspace{-2.4cm}
\begin{minipage}[t]{0.35 \linewidth}
\begin{center}
\mbox{\epsfysize=\hsize\epsffile{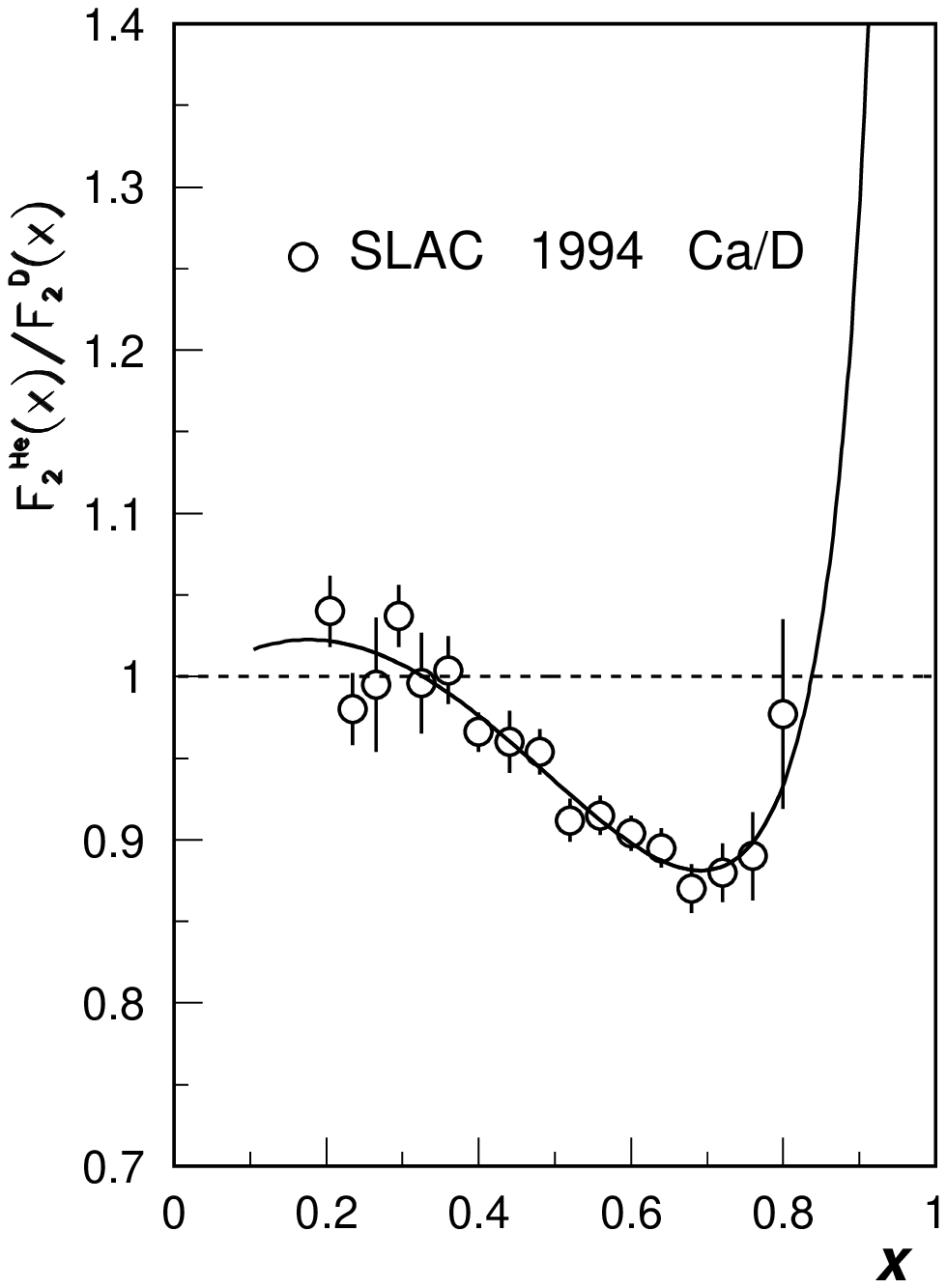}}
\end{center}
\end{minipage}
\hspace{-2.4cm}
\begin{minipage}[t]{0.35 \linewidth}
\begin{center}
\mbox{\epsfysize=\hsize\epsffile{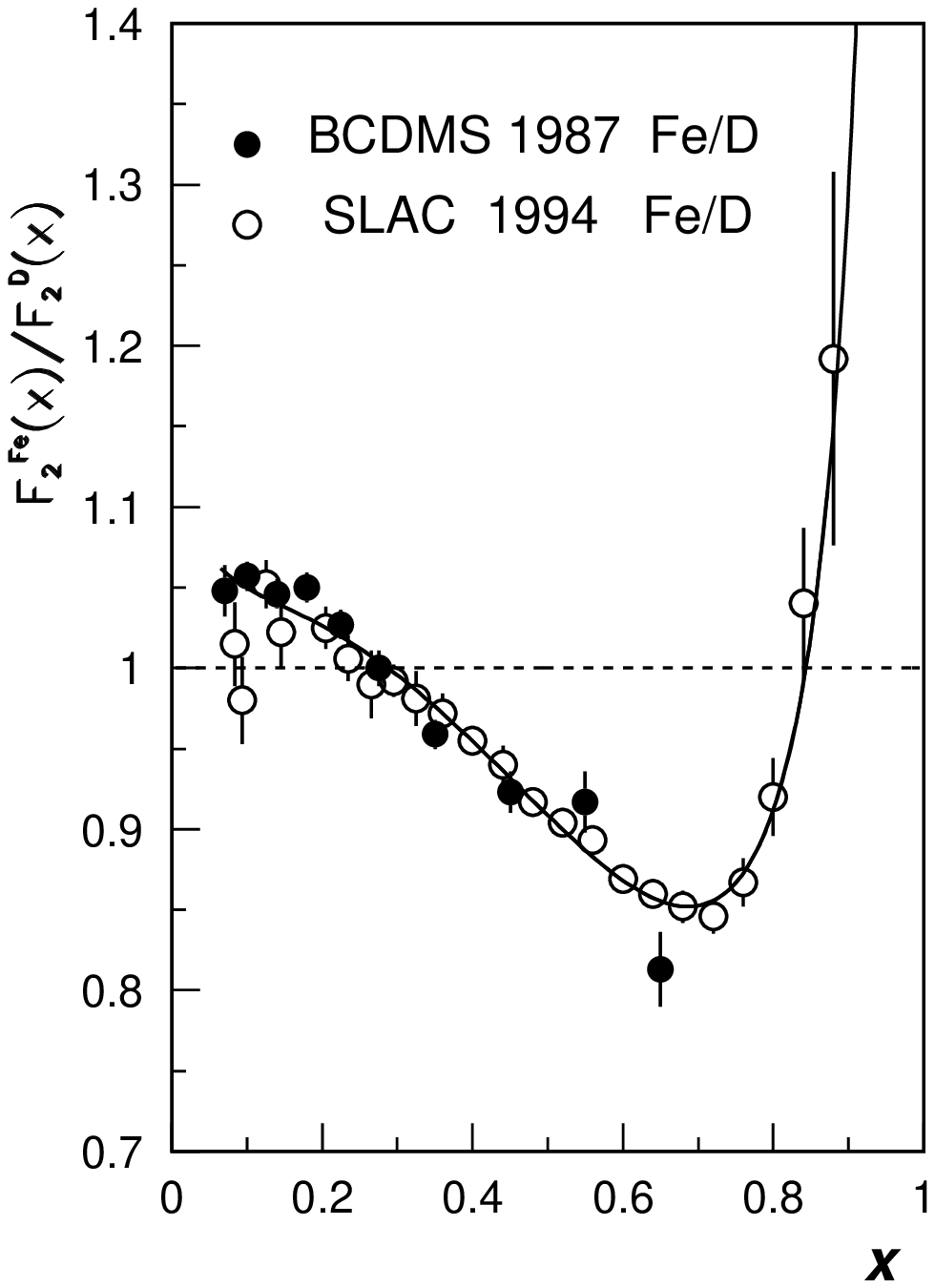}}
\end{center}
\end{minipage}

\caption{
 The results of the fit with Eq.(3) of the $F_2^A/F_2^{\rm D}$
measured on $^4$He [6], Be [6], Ca~[6] and Fe~[6,7]
 $r^A(x)$ in the high-$x$ range.} 
\end{figure}

Though Eq. (\ref{smir1}) represents only rough approximation of
nuclear effects when $x \to$ 1, it provides good description of the
data  available in the high-$x$ range (c.f. Fig. 2) and thus serves our purpose
of quantitative determination of the distortions of the nucleon structure function. 
 The obtained parameters $m_i$, which are displayed in Fig. 3,
increase from their minimum value $m_i$(He) at $A$ = 4 
to $m_i(A)$ $\approx$ 3$m_i$(He) for $A>$ 40,
indicating that  distortions in heavy nuclei are
independent of the size of the nucleus.

\begin{figure}[p]
\begin{center}
\mbox{\epsfxsize=0.7\hsize\epsffile{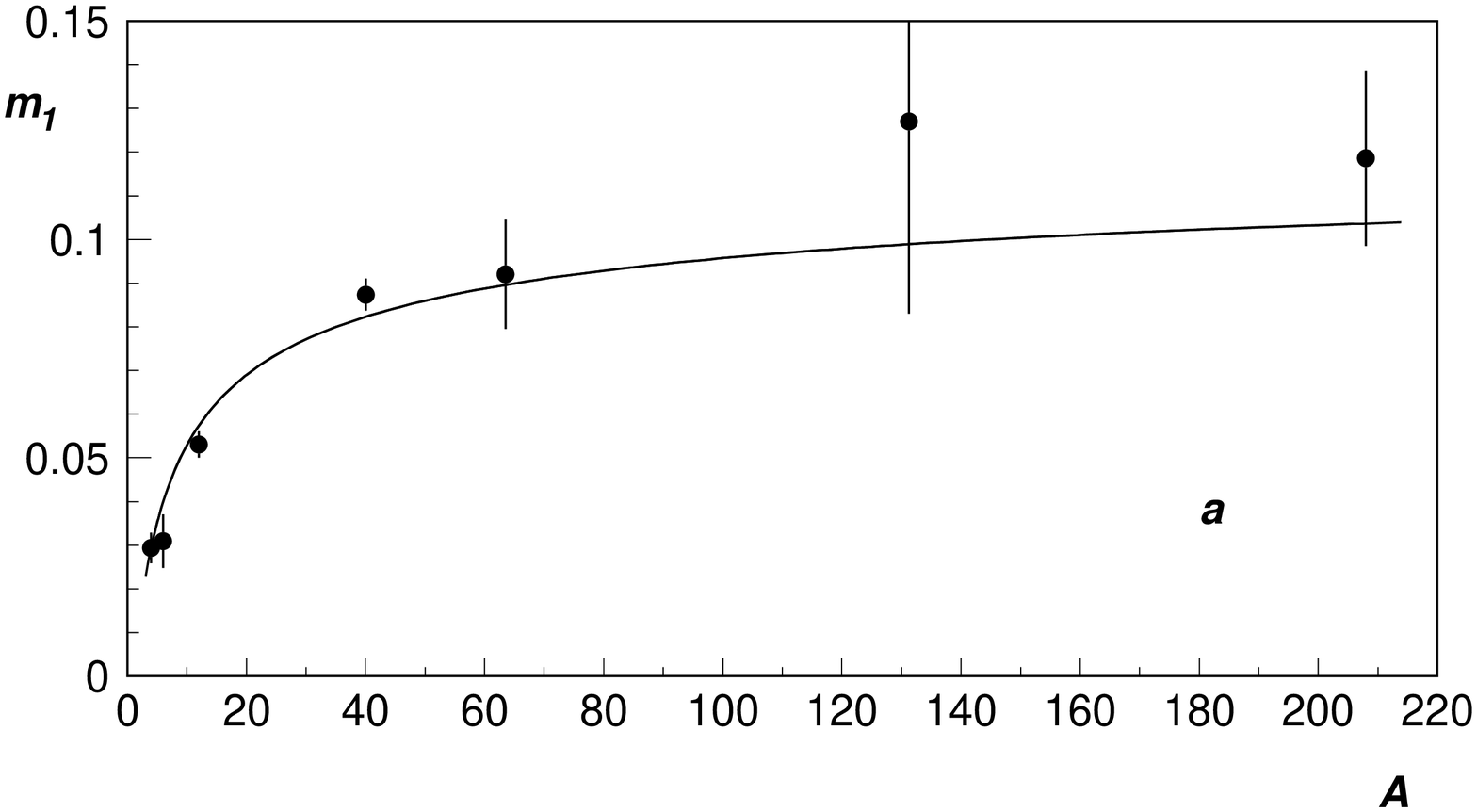}}
\end{center}
\vspace{-1.2cm}

\begin{minipage}[t]{0.38\linewidth}
\begin{center}
\mbox{\epsfysize=\hsize\epsffile{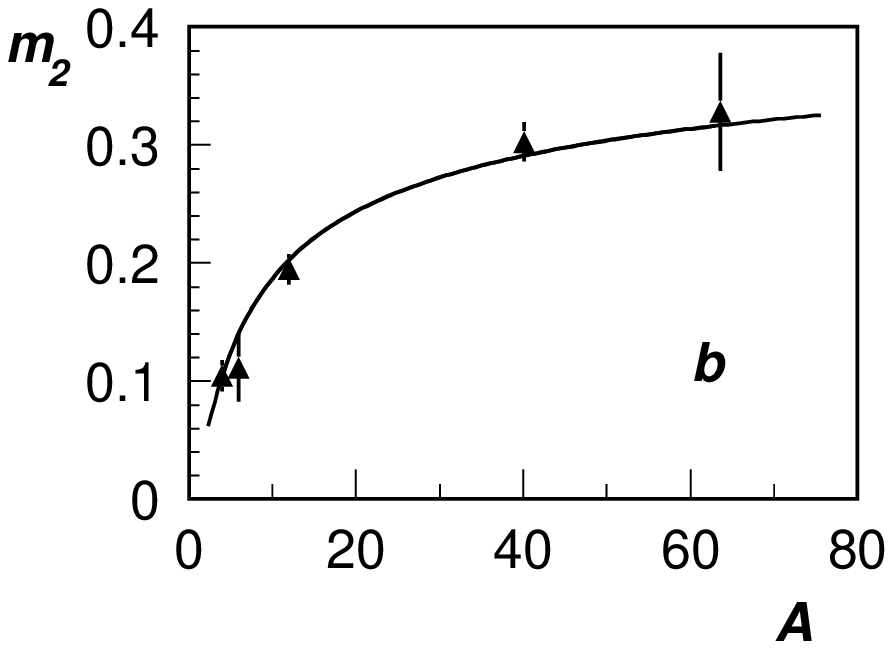}}
\end{center}
\end{minipage}
\hfill
\begin{minipage}[t]{0.38\linewidth}
\begin{center}
\mbox{\epsfysize=\hsize\epsffile{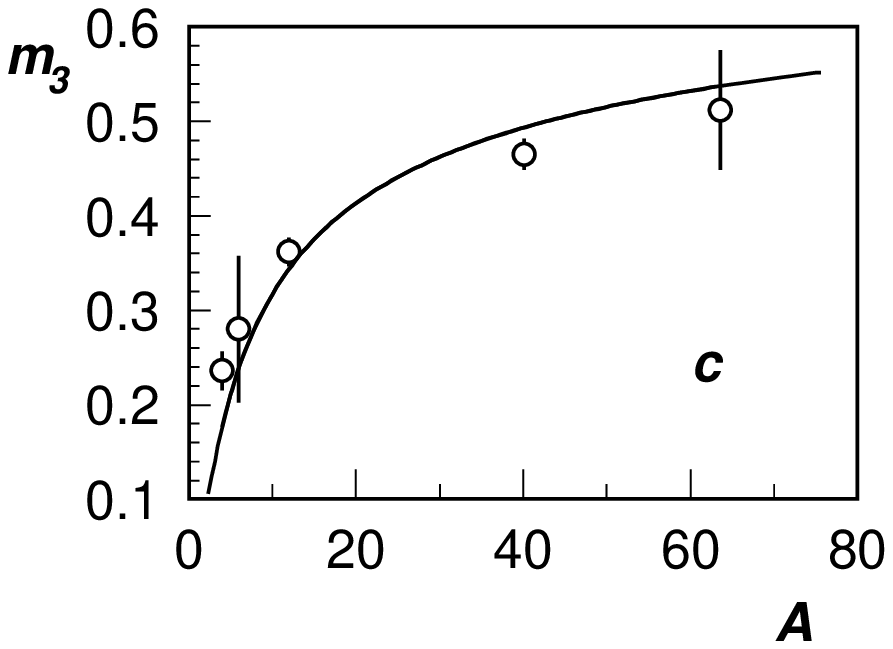}}
\end{center}
\end{minipage}
\vspace{-1cm}

\begin{center}
\mbox{\epsfxsize=0.7\hsize\epsffile{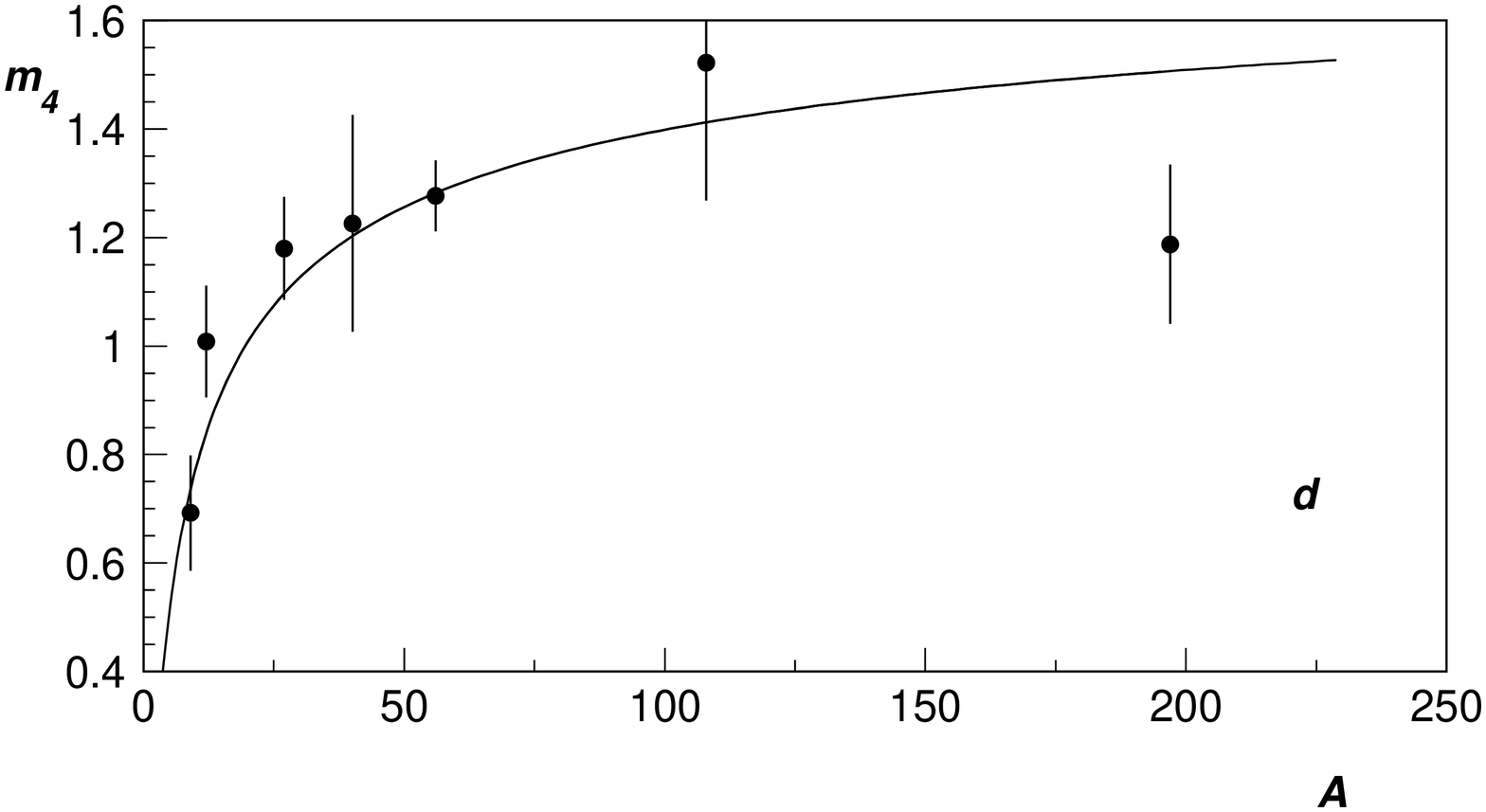}}
\end{center}
\caption{ 
 The parameters $m_i$, $i$= 1 -- 4,
 determined in the  regions of nuclear shadowing (a),
 anti-shadowing (b),  EMC effect  (c) and in the high $x$-range (d). 
Full lines show a variation in nuclear density given by the Woods--Saxon
 potential, with parameters fixed from the data on elastic
 electron--nucleus scattering.}
\end{figure}

The  parameters $m_i$  vary similarly with  $A$ 
in all four intervals in which the distortions
were depicted. 
The points in Fig. 3 are approximated by the following
equation:
\begin{equation}
  m_i (A) = N_i \Biggl( 1 - \frac{A_{\rm S}}{A} \Biggr)  ~.
\label{smbar}
\end{equation}
This coincides, except  for the normalization parameter $N_i$,
with the factor $\delta (A)$ suggested in Ref.~\cite{barsh}
for explaining the $A$ dependence of the EMC effect
by variation of the  nuclear surface-to-volume ratio with $A$.
The number of nucleons $A_{\rm S}$ at the nuclear surface was
obtained in Ref.~\cite{barsh}
using a Woods--Saxon potential with parameters taken from
Ref.~\cite{bohr}.

The four lines in Fig. 3, a, b, c and d, differ only in the 
normalization factor $N_i$.  The observed similarity in variations of  $m_i$
 give evidence for a {\em universal} $A$ dependence of the distortion magnitudes
$m_i$ of the nucleon structure function in all four regions.  This universality  can
 be expressed  in terms of the relative  distortions, measured in
nuclei  $A_1$ and $A_2$ with the following relation:
\begin{equation}
 {m_1(A_2) \over m_1(A_1) } ~=~ 
{m_2(A_2) \over m_2(A_1) } ~=~ {m_3(A_2) \over m_3(A_1) } ~=~
{m_4(A_2) \over m_4(A_1) } ~.
\label{simil}
\end{equation}

One can as well define the value of structure
function distortion in units of that measured in the helium nucleus, 
$s_{\rm h}$ = $m_i(A)$ / $m_i$(He).
 By definition, $s_{\rm h}$ = 1 for $A$ = 4, and,
as follows from the obtained numerical values of $m_i$,
$s_{\rm h}$ increases with $A$ to $\sim$~3 for heavy nuclei, independent
 of $x$. Eqs. (\ref{smir}) and (\ref{smbar}) provide  the approximation
of $F_2^A(x) / F_2^{\rm D}(x)$  in two dimensions displayed in Fig. 4 
 valid in the range of the present analysis, $A \geq$ 4, 10$^{-3}$ $<x <$ 0.7.

 \begin{figure}[p]
%
 

\parbox[m]{8cm}{\epsfig{bbllx=5mm,bblly=25mm,bburx=100mm,bbury=225mm,%
      file=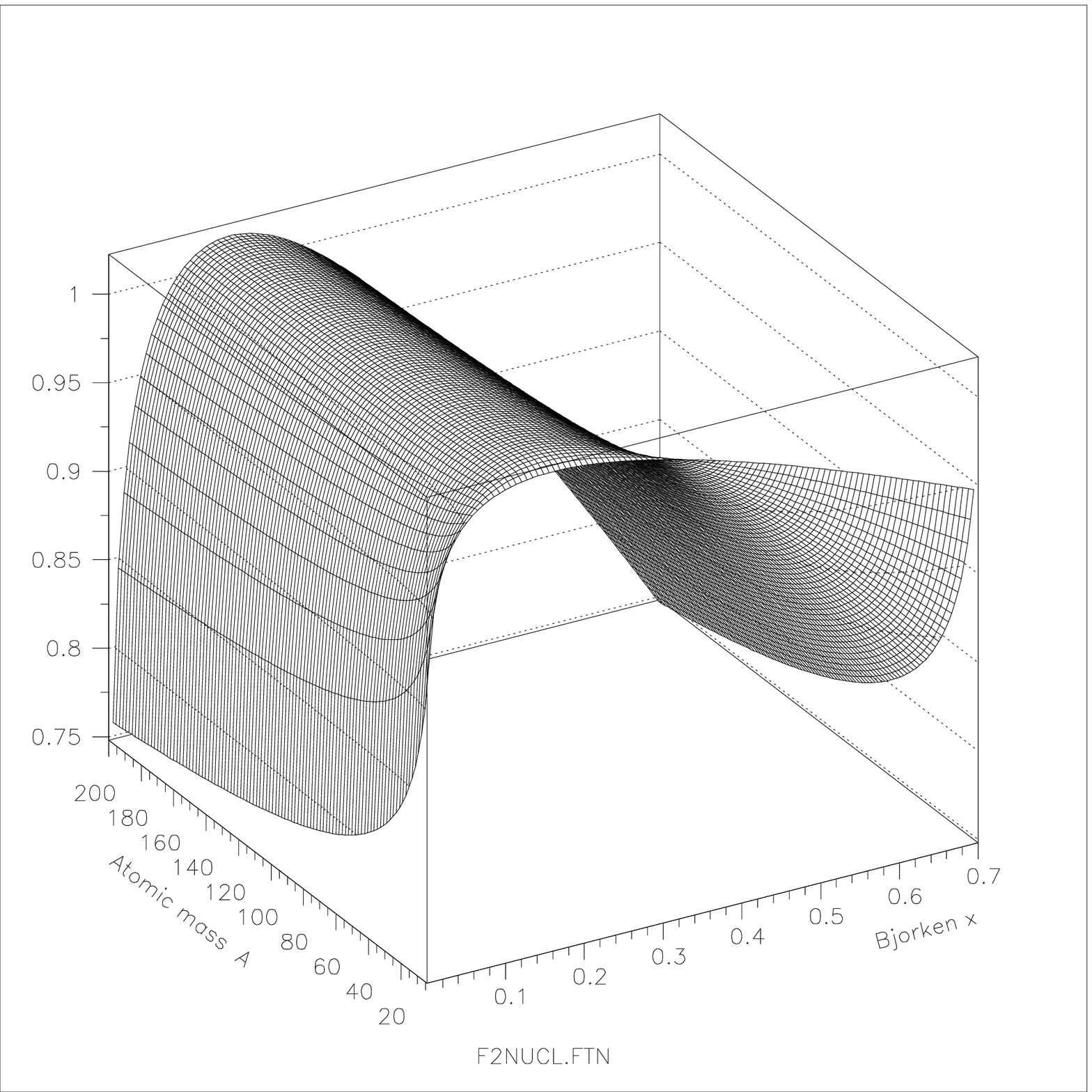,height=6cm,width=6.5cm}}

\vspace{2.5cm}

\caption
{  Approximation of the  $F_2^A(x) / F_2^{\rm D}(x)$ as a function
of atomic mass $A$ and $x$ in the range $A \geq$ 4, 10$^{-3}$ $<x <$ 0.7~.}
\end{figure}

We suggest that modifications to the
parton distributions of the nucleon bound in a nucleus
 evolve as a function of atomic mass $A$
in two stages. In the first stage, the distributions of
partons belonging to the
lightest nuclei, 2 $< A \leq$ 4, are modified drastically
 compared to those of a free nucleon, thus distorting
the  structure function $F_2(x)$.
These distortions, which can be observed in a $^4$He 
nucleus as  a characteristic oscillation of $r^A$
around the line  $r^A$ = 1,  remain frozen in shape
in the second stage of distortions, which occur in nuclei with
mass  $A >$ 4. 
In contrast to the first stage, in the second 
 there is no restructuring of parton distributions,
 which can change the shape
of the oscillation described by Eq. (\ref{smir}). Instead, the
distortions increase in magnitude throughout the
entire $x$ range, following the functional form (\ref{smbar}).
 
There  are evidently two different mechanisms behind this
picture, which we denote as {\em hard} or {\em soft}
distortions, depending on whether $A \leq$ 4 or $A >$ 4.
Quantitatively, this can be expressed with the parameter $s_{\rm h}$,
which rapidly changes in the range of hard distortions,
from 0 to 1 ($\Delta A$ = 2), and only slowly in the range of
soft distortions, from 1 to $\sim3$ ($\Delta A \approx$ 200).
A particular case of the hard distortion mechanism, which works at
$A$ = 4,  has been
considered in Refs.~\cite{fk83,kon84}, in which EMC effect was
explained by the 12-quark  structure of nuclei. The dynamical model  for the EMC
effect, in which  gluons and exchanged quarks (antiquarks) 
are responsible for the interaction between three (or four) nucleons,
has been suggested in Refs.~\cite{barsh,brein}.

In terms of the two-mechanism  model, the experimental 
observations can be interpreted as follows:  hard distortions are saturated at $A$ = 4,
which can be understood if modifications of parton distributions in the 
nuclear environment are closely related to short-range nuclear forces. 
In this picture  functional form of $r^A(x)$ and, therefore, positions of the
three cross-over points with the line $r^A$ = 1  should be different
when they are obtained in  $^3$He and $^4$He nuclei. Before such data are
available one can not  exclude the possibility that the saturation  
is  reached at $A$ = 3.

In summary,
we have shown that the recent data on the DIS of electrons
and muons off nuclei bring new evidence for the universality of
the $x$ and $A$ dependence of distortions of a free-nucleon
structure function, $F_2(x)$, by a nuclear medium, when $A \geq$ 4.
Such  universality  imply  that hard distortions of parton distributions
are saturated at $A$ = 4 (or even at $A$ = 3) and that
the observed differences between the DIS 
cross-sections for nuclei with masses $A_1$, $A_2 \geq$ 4
are due to soft distortions.  
The latter are similar in the entire $x$-range
and vary from 1 in $^4$He to $\sim$3 in $^{207}$Pb. They  can
be well understood as a nuclear density effect if the
surface nucleons are excluded from consideration.


 

\begin{thebibliography}{999}
\vspace{0.5cm}

\baselineskip=0.666\baselineskip
\vspace{-0.3cm}
\bibitem{sm94}  G.I. Smirnov, Yad. Fiz., {\bf 58}, No. 9 (1995) 1712;
Phys. At. Nucl. (Engl. Transl.), {\bf 58}, No. 9 (1995) 1613;
hep--ph--9502368.
\vspace{-0.3cm}
\bibitem{sm95}  G.I. Smirnov, Phys. Lett. {\bf B 364} (1995) 87; 
hep--ph--9512204.
\vspace{-0.3cm}
\bibitem{ama95} NMC, P. Amaudruz et al., Nucl. Phys. {\bf B 441} (1995) 3.
\vspace{-0.3cm}
\bibitem{arn95} NMC, M. Arneodo et al., Nucl. Phys. {\bf B 441} (1995) 12.
\vspace{-0.3cm}
\bibitem{ad95}  E665, M.R. Adams et al., Z. Phys. {\bf C 67} (1995) 403;
hep--ex--9505006. 
\vspace{-0.3cm}
\bibitem{gomez} SLAC, J. Gomez et al., Phys. Rev. {\bf D 49} (1994) 4348.
\vspace{-0.3cm}
\bibitem{bari} BCDMS, G. Bari et al., Phys. Lett. {\bf B 163} (1985) 282;\\
  A.C.Benvenuti et al., Phys. Lett. {\bf B 189} (1987) 483.
\vspace{-0.3cm}
\bibitem{marti} A.D. Martin, W.J. Stirling and R.G. Roberts, 
 Phys. Rev. {\bf D 50} (1994) 6734.
\vspace{-0.3cm}
\bibitem{rep160} L. Frankfurt and M. Strikman, Nucl. Phys.
 {\bf B 250} (1985) 143;\\
 L. Frankfurt and M. Strikman, Phys. Rep. {\bf 160} (1988) 235.
\vspace{-0.3cm}
\bibitem{xe92}  E665, M.R. Adams et al., Phys. Rev. Lett.
 {\bf 68} (1992) 3266.
\vspace{-0.3cm}
\bibitem{copper} EMC, J. Ashman et al., Z. Phys. {\bf C 57} (1993) 211.
\vspace{-0.3cm}
\bibitem{barsh}  S. Barshay, Z. Phys. {\bf C 27} (1985) 443;\\
 S. Barshay and D. Rein, Z. Phys. {\bf C 46} (1990) 215.
\vspace{-0.3cm}
\bibitem{bohr} A. Bohr and M. Mottelson, {\em Nuclear Structure},
 W.A.Benjamin, Inc., New York, 1969.
\vspace{-0.3cm}
\bibitem{fk83} H. Faissner and B.R. Kim, Phys. Lett. {\bf B 130} (1983) 321;\\
H. Faissner, B.R. Kim and H. Reithler, Phys. Rev. {\bf D 30} (1984) 900.
\vspace{-0.3cm}
\bibitem{kon84} L.A. Kondratyuk and M.Zh Shmatikov, Pis'ma Zh.
Eksp. Teor. Fiz. {\bf 39} (1984) 324 (JETP Lett. {\bf 39} (1984) 389);\\
Yad. Fiz. {\bf 41} (1985) 498 (Sov. J. Nucl. Phys. {\bf 41} (1985) 317).
\vspace{-0.3cm}
\bibitem{brein}  S. Barshay and D. Rein, Particle World, {\bf 4}  No 2 (1994) 3.

\end{thebibliography}
\end{document}